# OpenSQUID: a flexible open-source software framework for the control of SQUID electronics

Felix T. Jaeckel, Randy J. Lafler, and S. T. P. Boyd, Member

*Abstract*—Commercially available computer-controlled SQUID electronics are usually delivered with software providing a basic user interface for adjustment of SQUID tuning parameters, such as bias current, flux offset, and feedback loop settings. However, in a research context it would often be useful to be able to modify this code and/or to have full control over all these parameters from researcher-written software. In the case of the STAR Cryoelectronics PCI/PFL family of SQUID control electronics, the supplied software contains modules for automatic tuning and noise characterization, but does not provide an interface for user code. On the other hand, the Magnicon SQUIDViewer software package includes a public application programming interface (API), but lacks auto-tuning and noise characterization features. To overcome these and other limitations, we are developing an "open-source" framework for controlling SQUID electronics which should provide maximal interoperability with user software, a unified user interface for electronics from different manufacturers, and a flexible platform for the rapid development of customized SQUID auto-tuning and other advanced features. We have completed a first implementation for the STAR Cryoelectronics hardware and have made the source code for this ongoing project available to the research community on SourceForge (http://opensquid.sourceforge.net) under the GNU public license.

## I. INTRODUCTION

THE READOUT of superconducting quantum interference devices (SQUIDs) requires bias circuitry, low noise electronic amplifiers, and often feedback loops for linearization. A variety of integrated SQUID electronics have been developed [1-6] and several such systems are commercially available [7-11]. These commercial systems are typically supplied with a PC-based software package giving control, via a graphical user interface (GUI) [12], of the SQUID tuning parameters, like bias current, flux, and modulation. Through our experience with both the STAR Cryoelectronics (STARCryo) [7] and Magnicon [8] products,

Manuscript received October 9, 2012. Support for this work was provided by the *U.S. Department of Energy*, the *Defense Threat Reduction Agency*, and the *National Science Foundation*.

Felix T. Jaeckel (corresponding author) is with the Department of Physics and Astronomy, University of New Mexico, Albuquerque, NM 87131 USA (phone: 505-620-4876; fax: 505-277-1520; e-mail: jaeckel@unm.edu).

Randy J. Lafler is with the Department of Physics and Astronomy, University of New Mexico, Albuquerque, NM 87131 USA (e-mail: rlafler@unm.edu).

S. T. P. Boyd is with the Department of Physics and Astronomy, University of New Mexico, Albuquerque, NM 87131 USA (e-mail: stpboyd@unm.edu).

we have found that these closed-source software packages suffer from several limitations that can be troublesome in a research context.

In the case of the STARCryo PCS-10X software, advanced features like auto-tuning, calibration, and noise measurements are available when the STARCryo SQUID electronics is supplemented with PC-based data-acquisition hardware from National Instruments [13]. However, some very useful basic features like automatic flux-reset and flux-counting, or even simultaneous operation of modulation and two-stage array readouts are not supported. On the other hand, the Magnicon SQUIDViewer software includes automatic flux-reset and flux-counting, but offers no auto-tuning, feedback calibration, or noise recording functionality.

One area where both software packages display significant limitations is in their ability to integrate with user-written software. Software integration is an important requirement for the development of automated measurement systems, where SQUIDs may need to be reset or retuned frequently due to automated changes in experimental parameters, such as operating temperature or excitation. The STARCryo software provides no support for software integration. The Magnicon software does provide a public application programming interface (API), but it is then still up to the user to duplicate the functionality of the vendor-supplied, closed-source GUI.

Furthermore, many applications of current interest, especially those where a large number of SQUIDs are in use, would benefit from additional software capabilities, such as single-click documentation of tuning parameters and noise measurements. When SQUID sensors need to be tuned remotely (e.g. in geomagnetic networks), the software should provide a way to visualize the transfer function live within the same GUI. Where the hardware supports it, the software should also allow simultaneous operation of SQUIDs with different readout schemes (i.e. modulation or direct readout), a capability which is absent from the STARCryo software package.

To overcome these limitations of the vendor-supplied software packages, we have begun development of an open-source software SQUID control framework with an object-oriented, modular architecture designed to support the aforementioned features. At the time of writing, we have completed software for the STARCryo PCI-1000/PFL-100/PFL-102 SQUID electronics that provides the full functionality of the vendor-supplied software plus the following new features: 1) simultaneous operation of PFL-100 and PFL-102 from the same PCI-1000 control electronics,

2) automatic software flux reset and reset counting, 3) automated tuning, feedback calibration, and noise measurements of the PFL-102 STARCryo two-stage SQUID electronics (the vendor-supplied software has full functionality only for their PFL-100 modulation electronics), 4) single-click documentation of tuning parameters, $V$-$\phi$ curves, and noise measurements, and 5) open-source code exposing all functionality and data of the SQUID control software to user-written software. This new software is now the standard software for control and readout of STARCryo SQUID electronics in our lab.

In the following sections, we describe the development process, the software architecture, and the new capabilities that have been implemented.

A note on nomenclature: a number of schemes have been developed for reading out SQUIDs. Because this report focuses on software developed for the STARCryo PCI-1000/PFL-100/PFL-102 SQUID electronics, we will sometimes use "modulation" as shorthand for the PFL-100 flux-modulation flux-locked-loop (FLL) electronics reading out a single dc SQUID, and "array" or "two-stage" as shorthand for the PFL-102 two-stage SQUID FLL electronics using directly-coupled readout [14][15], with a single dc SQUID as the first stage, and a STARCryo SQUID array as the second stage [16].

## II. DEVELOPMENT AND PROGRAMMING LANGUAGE

We chose to focus initial software development on the STARCryo PCI-1000/PFL-100/PFL-102 SQUID electronics, which are used for the majority of our measurements. With the consent of Robin Cantor, president of STAR Cryoelectronics, we investigated the serial communication protocol used between the PC and the STARCryo SQUID electronics by the simple expedient of listening to it with a port sniffer. An initial prototype of the new control software was developed in LabView [13] and used to validate our implementation of the communication protocol.

With the LabView prototype we were able to reproduce the basic functionality of the vendor-supplied software. We were also able to implement the first new capability: simultaneous operation of both array and modulation flux-locked-loops (FLLs) on the same PCI-1000 unit. However, when we attempted to implement more complex algorithms needed for automated SQUID tuning we found the LabView graphical programming approach to be inefficient. Since we had already decided separately to phase out the use of LabView in our lab due to its limitations for complex projects, we decided to re-implement the SQUID control software in a text-based procedural language before continuing development.

We considered several alternative programming environments. Matlab [17] has good support for data analysis, plotting, and GUIs. Toolboxes for instrument control and data acquisition are also available. However, multithreaded or multi-process program execution is difficult to achieve and limits code modularity. Thus Matlab would not support our preferred approach for instrument control, which is a large number of simple programs running simultaneously and communicating with each other.

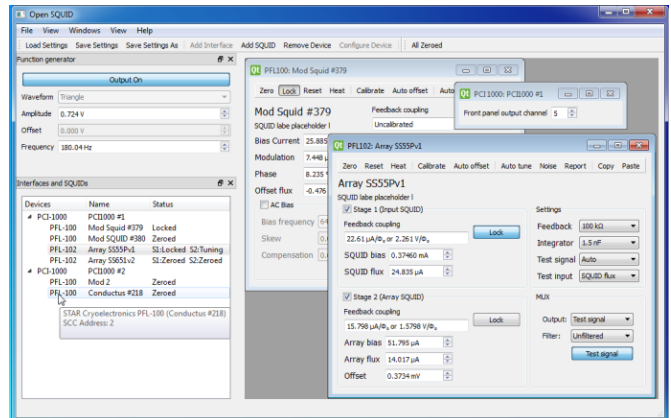

Fig. 1. Screenshot of the main window of the GUI, showing active controls for both modulation (PFL-100) and two-stage (PFL-102) readout.

In the case of C++, sophisticated mechanisms and libraries exist to fulfill all our requirements, but we were concerned that the learning curve for C++ may be too steep for most students in the lab.

Python [18] presents a compromise that has become popular in research labs for a number of reasons: ease of use, availability of numerical tools comparable to Matlab through the NumPy [19] and SciPy [20] projects, good performance for an interpreted language, the possibility to interface with existing C and C++ code and binary libraries, as well as the availability of a large set of supporting libraries, including good support for instrument control via PyVisa [21] and inter-process communications via ZeroMQ [22]. Since a user-friendly GUI is a primary requirement for us, we decided to make use of the popular Qt libraries [23], which provide a comprehensive, extensible framework of considerable sophistication and flexibility. Qt library functionality is readily leveraged within Python through PyQt [24] and PyQwt [25]. Note that this entire programming "stack" has been aggregated into a single easily-installed package by the Python(x,y) project [26] to simplify Python adoption by the scientific and engineering community.

## III. ARCHITECTURE

### A. Software

To facilitate the eventual integration of SQUID electronics from multiple vendors, a modular approach was chosen. The object-oriented paradigm is a good match for the representation of real-world objects in software. The STARCryo PCI-1000/PFL-100/PFL-102 system was abstracted into several components. All components of the system communicate through the computer interface provided by the PCI-1000 class. With this "separation of responsibilities" network-transparent remote operation can easily be supported in the future. The PCI-1000 function generator and multiplexer, although physically contained in the same box, are represented as separate objects. This modularity makes the code flexible enough to be easily extended for other hardware sets, where the components of the system differ. For example, owners of the PCI-100 (which does not have an integrated function generator) would only need to program an interface class for a function generator of their choice to use the software with an otherwise undiminished feature set. Similarly, we strive to keep the graphical interface modular and separate from core functionality, so that users can compose their own GUI based on their specific needs with a minimum amount of effort. A screenshot of the supplied default GUI is shown in Fig. 1.

TABLE I
INSTRUMENTATION USED IN SOFTWARE DEVELOPMENT

| Category | Vendor | Model |
|---|---|---|
| Data Acquisition | National Instruments | USB-4431 |
| Oscilloscope | Agilent | InfiniiVision DSO-X 2004A |

TABLE II
SUPPLEMENTAL INSTRUMENTATION REQUIRED FOR FUNCTION

| Functionality | Minimum Instrumentation Required |
|---|---|
| Adjust settings manually with GUI | None |
| Auto-tune/calibrate/noise spectra with modulation electronics | 100 kS/s, ±10 V, ≥16 bits 2-channel simultaneous sampling DAQ |
| Auto-tune/calibrate/noise spectra with STARCryo two-stage SQUID amplifier electronics | above plus digital storage oscilloscope or 5 MS/s 2-channel digitizer |

The Qt signal-slot mechanism is used to couple GUI components to the underlying hardware objects, ensuring consistency in the display and permitting the use of multithreading to prevent unresponsive user interfaces.

*B. Accommodating Instrumentation Variation*

Implementing advanced functionality in SQUID control software requires additional instrumentation to support the SQUID electronics. For example, minimal auto-tuning with the STARCryo modulation electronics requires the ability to measure and record $V$-$\phi$ curves. The necessity to have supporting instruments creates an unavoidable complication for any software project of this type, because other researchers may not have access to the same instruments. We have taken steps in both the software architecture and in the choice of development hardware to minimize this problem. The instrumentation used for the software development is fairly generic, as indicated in Table I.

In the software architecture, setup screens are provided for each SQUID channel to configure the choice of data-acquisition instrument and oscilloscope. If the user's instrumentation is not compatible with the development hardware shown in Table I, it is easy to create and substitute a new Python class to support that instrumentation.

Table II shows the hardware requirements to achieve a given functionality in the present version of the code. If a user wants to evaluate the new software, it can be installed in a system with no supporting instrumentation and it will still provide the basic functionality to manually tune and operate SQUIDs.

## IV. NEW FEATURES

The new software provides the full functionality of the vendor-supplied software plus several useful new features, which are described in more detail in this section.

*A. Simultaneous operation of modulation and array FLLs*

Simultaneous use of both PFL-100 modulation electronics and PFL-102 two-stage electronics from the same PCI-1000 control electronics is now supported. Interface windows for both types of PFL are shown in the screenshot of the main GUI window in Fig. 1. This capability can significantly reduce the hardware complexity and expense necessary to operate SQUID readouts of both types.

*B. Software-based automatic flux reset and reset counting*

The STARCryo SQUID electronics and software provide no support for hardware or software-based automatic FLL reset and reset counting. Although a hardware-based solution would yield faster performance, a software-based solution can still be very useful, and has been implemented. In contrast to the Magnicon approach, it can easily be customized for user applications due to its open-source nature.

*C. Auto-tuning for PFL-102 electronics*

Auto-tuning, feedback calibration, and integrated noise spectra have now been implemented for the PFL-102 two-stage electronics. The auto-tuning GUI for the PFL-102 is shown in Figs. 2 and 3, and the noise spectrum GUI is shown in Fig. 4. The auto-tuning procedure used in this initial implementation, summarized in Table III, is mostly based on the manual tuning procedure recommended in the STARCryo documentation, combined with a final step of noise spectrum measurement and some minor added steps to ensure good results under automation.

However, note that the STARCryo documentation recommends using single-$\phi_0$ resets of the FLL to determine the feedback calibration (in units of $V/\phi_0$) for the cascaded two-stage amplifier. That approach can yield incorrect results for the PFL-102 because of the complex form of the $V$-$\phi$ curve

TABLE III
AUTO-TUNING STEPS FOR PFL-102 WITH INPUT SQUID + SQUID ARRAY

| Activity | Excitation Applied to | FLL Lock |
|---|---|---|
| adjust array bias current to maximize peak-to-peak amplitude of array $V$-$\phi$ curve | array input | unlocked |
| adjust array offset to symmetrize maximum and minimum of array $V$-$\phi$ curve about zero volts | array input | unlocked |
| obtain array feedback calibration from single-$\phi_0$ FLL resets | none | array |
| adjust SQUID bias current to maximize peak-to-peak amplitude of SQUID $V$-$\phi$ curve | SQUID feedback | array |
| adjust array flux to symmetrize maximum and minimum of SQUID $V$-$\phi$ curve about zero volts | SQUID feedback | array |
| obtain SQUID feedback calibration from autocorrelation of SQUID $V$-$\phi$ curve. | SQUID feedback | array |
| obtain noise spectrum for SQUID array | none | SQUID |

"SQUID" in this table indicates the first stage input SQUID. "Array" indicates the second stage SQUID array. See reference [16] or STARCryo PFL-102 documentation for the two-stage SQUID amplifier circuit diagram.

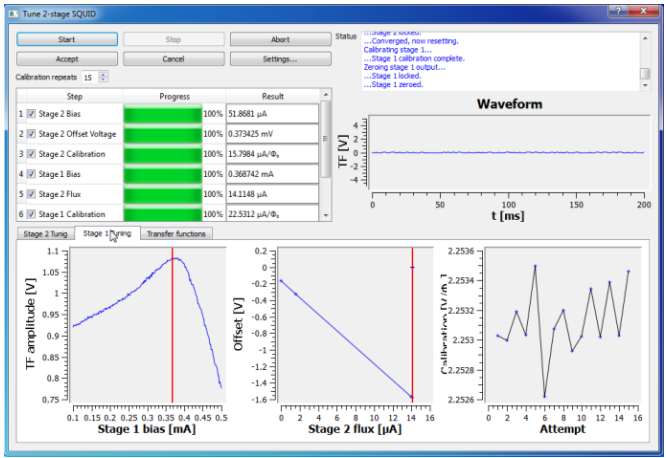

Fig. 2. Screenshot of the auto-tuning interface for STARCryo PFL-102 two-stage SQUID electronics. While tuning is in progress the "waveform" graph shows the $V$-$\phi$ curve in real time, but this is difficult to capture in a screenshot. The three graphs in the tabbed lower panel show tuning results for the input SQUID (stage 1). Left: Amplitude of $V$-$\phi$ curve versus bias current. Middle: Mean value of $V$-$\phi$ curve versus array (stage 2) flux offset. Right: Stage 1 feedback calibration determined by autocorrelation of the $V$-$\phi$ curve.

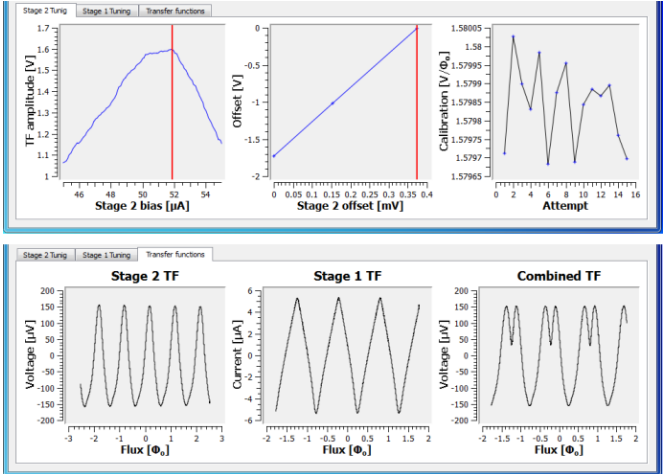

Fig. 3. Screenshots of the other two tabs of the lower panel in the PFL-102 tuning interface. Upper panel: SQUID array (stage 2) tuning results, see description in Fig. 2. Lower panel: The three $V$-$\phi$ curves (transfer functions) after completion of the auto-tuning procedure. Left: SQUID array. Middle: Input SQUID. Right: The combined $V$-$\phi$ curve.

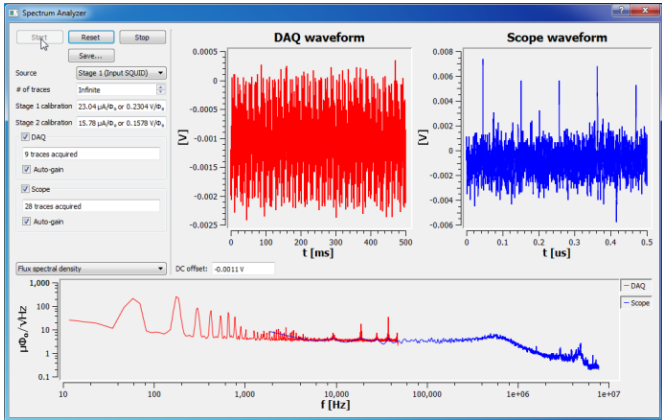

Fig. 4. Screenshot of noise spectrum interface, showing noise performance and bandwidth obtained with PFL-102 two-stage readout. To obtain coverage of the full readout bandwidth, measurements using the NI DAQ device are combined with measurements using a digital oscilloscope. For the DAQ device, aliasing of high-frequency signals is prevented by the oversampling of the delta-sigma analog-to-digital converter and integrated low-pass filters.

when the input SQUID and the SQUID array are cascaded [27]. Instead, our auto-tuning scheme for the PFL-102 deduces the 1st stage feedback calibration from the auto-correlation of the input SQUID's $V$-$\phi$ curve taken over multiple periods. The entire auto-tuning procedure typically completes in less than 30 seconds.

### D. Single-Click Recordkeeping

Keeping detailed records of tuning and performance characterization of new SQUID devices with vendor-supplied software is time consuming: one must manually record the tuning parameters, choices for feedback resistance and integrator capacitance, feedback calibration, noise spectra and traces of the $V$-$\phi$ curve. The new software simplifies documentation by implementing single-click export of all information for a given device to a single Matlab MAT-file [28], from which it is easy and quick to extract data and make plots. The new software also improves on our existing documentation procedure by including all of the detailed results of the tuning procedures, *e.g.* the variation of the peak-to-peak amplitude of the $V$-$\phi$ curve versus bias current. Again, the software's open-source nature allows for full customization based on user-specific requirements.

## V. FUTURE WORK

Now that the large structures of auto-tuning and noise measurement are in place and tested it will be easy to add evolutionary improvements such as more sophisticated optimization goals for the auto-tuning, *i.e.* maximizing $dV/d\phi$ or minimizing measured noise. It should also be straightforward to modify the software to support the STARCryo PCI-100 single-readout control electronics.

The next major effort for the project will be to integrate support for the Magnicon XXF-1 SQUID electronics. In support of this, Magnicon has provided us with a detailed description of their low-level communications protocol [29].

## VI. CONCLUSION

We have written new control software for the STARCryo PCI-1000/PFL-100/PFL-102 SQUID electronics that provides the full functionality of the vendor-supplied software plus several useful new features as described above. The software is structured so that it can be extended to support SQUID electronics and supplemental instrumentation from other vendors. This software may be useful to other researchers, so it has been made available to the community as the initial release of the "OpenSQUID" open-source software project at http://opensquid.sourceforge.net under the GNU Public License (GPL). We invite code contributions from the research community and vendors.

## ACKNOWLEDGMENT

The work reported here was performed to support multiple projects funded by the Defense Threat Reduction Agency, the US Department of Energy (NA-22), and the National Science Foundation.